# Identifying Social Satisfaction from Social Media

Shuotian Bai, Rui Gao, Bibo Hao, Sha Yuan, and Tingshao Zhu*
School of Computer and Control Engineer, Institute of Psychology, CAS, Beijing

**Abstract.** We demonstrate the critical need to identify social situation and instability factors by acquiring public social satisfaction in this research. However, subject to the large amount of manual work cost in subject recruitment and data processing, conventional self-reported method cannot be implemented in real time or applied in large scale investigation. To solve the problem, this paper proposed an approach to predict users' social satisfaction, especially for the economy-related satisfaction based on users' social media records. We recruited 2,018 Cantonese active participants from each city in Guangdong province according to the population distribution. Both behavioral and linguistic features of the participants are extracted from the online records of social media, i.e., Sina Weibo. Regression models are used to predict Sina Weibo users' social satisfaction. Furthermore, we consult the economic indexes of Guangdong in 2012, and calculate the correlations between these indexes and the predicted social satisfaction. Results indicate that social satisfaction can be significantly expressed by specific social media features; local economy satisfaction has significant positive correlations with several local economy indexes, which supports that it is reliable to predict social satisfaction from social media.

## Introduction

### Deficiencies of Traditional Social Satisfaction Survey

As human society is made up of individuals, the satisfaction of its members plays a key role while measuring social situation. By assessing and understanding the social satisfaction of the public in large scale investigation is the most important to identify social situation and instability factors.[1,2]

For most of such investigations, the investigators manage to interview the participants in household and let them complete self-report inventories to assess their satisfaction.[3,4,5] Although self-report is the widely used method to assess the participant's satisfaction, it still has several deficiencies in practice.[6] First, it is hard to recruit a large number of participants. Considering personal privacy, people commonly contradict the investigation and reject to take part in the survey. Second, social satisfaction is one kind of status variable which changes from time to time, the survey has to be conducted quite often to keep up with the changing tendency. Third, self-report method is quite time-consuming, which always takes several weeks or even months for post-processing. Coupled with the long time spent on recruiting participants, most large-scale investigations are carried out only once or twice a year.[6,7]

### Analysis of Psychological Variables in Social Media

With the development of information technology, social media (e.g., Twitter or Sina Weibo) becomes popular and widely used in daily life.[8] It is reported that the number of registered users of Chinese social media reaches 536 million at the end of 2013.[2] People spontaneously express themselves, which provides a large number of reliable data source for social science research. Under this trend, online behavior analysis is now adding its appeal for many researchers. With the use of Application Program Interfaces (APIs), the traceable online digital records of social media can be gained in bulk.

Previous research indicate that predicting psychological variables through social media features is feasible.[6,9,10] Michal et al.[11] showed that the easily accessible digital records of behavior in Facebook can be used to automatically and accurately predict a range of highly sensitive personal attributes such as personality, and happiness. Andrew et al.[12] analyzed 700 million words and phrases collected from the Facebook volunteers, who also took standard personality tests. They built personality classification models based on user's linguistic features. Gosling et al.[13] examined the manifestations of personality via SNS usage. They built a mapping between personality and Facebook online behaviors with 11 features, including the number of friends, weekly usage, etc. Based on both self-reported Facebook usage and observable profile information, they reported the correlation factor between personality and online behaviors. Li et al. managed to predict the active users' personality traits through micro-blogging behaviors of 547 Chinese active users. They extracted 845 micro-blogging behavioral features, and trained classification models utilizing Support Vector Machine (SVM) which differentiates participants with high and low scores on each dimension of the Big Five personality. The classification accuracy ranged from 84% to 92%. They also built regression models, which predicted participants' five dimensions of personality scores.

These works inspire us to predict users' social satisfaction by social media records. Until the end of 2013, the total count of Weibo users has reached 53.5 million in Canton (Guangdong province), China, which accounts for nearly 10% of all the Sina Weibo users.[23] During February, 2013, we recruited 2300 Cantonese Sina Weibo users and downloaded their Weibo records. Behavioral and linguistic features were extracted to build the social satisfaction prediction model. We trained the social satisfaction degree predicting model, which could be used to predict the public social satisfaction at any time interval. Furthermore, the variation tendencies of social satisfaction in the 21 Cantonese cities were demonstrated in this study. We then computed the correlation factors between our predicted social satisfaction of each city and its economy indexes from the official Guangdong Statistical Yearly Report[14], and found out that there exists significant relevance.

## Method

We made a survey with our experiment platform. Both online behavioral and textual linguistic features are extracted. We build the social satisfaction prediction model, and managed to dynamically describe the social satisfaction of the public in Guangdong province.

### Participants

In Canton, there are 21 cities (Chaozhou, Dongguan, etc., listed in Table 1). The invitations we send out were controlled by following the population distribution of Guangdong province. Table 1 shows the population distribution of Guangdong province, the number of invitations, and the valid participants recruited from each city.

Some participants did not meet our requirements since they are not active users[25]. Users who do not update any status in the last three months or less than 500 statuses in total are excluded. In our user samples, the number of statuses of each user published was 136 on average with a deviation of 788. Users who published less than 500 (approximately mean + 0.5*standard deviation) statuses were defined as inactive users.

Table 1. Sampling frame.

| City | Population (million) | Invitation count | Legal sample count |
|---|---|---|---|
| Chaozhou | 2.67 | 2600 | 147 |
| Dongguan | 8.22 | 7900 | 49 |
| Foshan | 7.20 | 6900 | 99 |
| Guangzhou | 12.70 | 12200 | 176 |
| Heyuan | 2.95 | 2800 | 116 |
| Huizhou | 4.60 | 4400 | 87 |
| Jiangmen | 4.45 | 1300 | 54 |
| Jieyang | 5.88 | 5600 | 6 |
| Maoming | 5.82 | 5600 | 125 |
| Meizhou | 4.24 | 4100 | 118 |
| Qingyuan | 3.70 | 3500 | 99 |
| Shantou | 5.39 | 5100 | 120 |
| Shanwei | 2.94 | 2800 | 129 |
| Shaoguan | 2.83 | 2700 | 110 |
| Shenzhen | 10.36 | 9900 | 69 |
| Yangjiang | 2.42 | 2300 | 99 |
| Yunfu | 2.37 | 2300 | 107 |
| Zhanjiang | 6.99 | 6700 | 100 |
| Zhaoqing | 3.92 | 3800 | 104 |
| Zhongshan | 3.12 | 3000 | 37 |
| Zhuhai | 1.56 | 1500 | 67 |

From February to April, 2013, 2658 Sina Weibo users participated in this study. The invited participants were instructed to log in "XinLiDiTu" with their Sina Weibo account and complete our satisfaction survey.

For each participant, our system recorded the exact time stamp while she/he submitted the answer of each question. We eliminated any participant who finished the inventory too quickly. Since the inventory consists of 15 Chinese characters on average for each question, we set the minimum response time (2s) on each question. That means if the time interval of two adjacent submissions is less than two seconds, this sample will be identified as an invalid one. All participants have been instructed to complete the survey seriously before they attend the experiment. Once they agreed to take part in the user study, their answers were carefully checked. A fee was paid only to the valid participants. Finally, the records of 2018 participants (892 females and 1126 males) with an average 22.34 years old (all are adults aged 18 or above) were remained.

**Survey Design**

We conducted the user survey on social satisfaction in Guangdong province, as it has the largest Weibo population in China. An online experiment platform named "XinLiDiTu" (http://ccpl.psych.ac.cn:10002/) was implemented, which was a web app of Sina Weibo. We invited participators by using the "@" function. The experiment used the *Urban and Rural Residents Social Attitudes Questionnaire*[27,28] which is a 13-question inventory measuring the satisfaction about life, income, social position, national economy, local economy and social justice.[15] Using Five-Likert questionnaire, the participants were instructed to complete the test and fill in the questionnaire through their Sina Weibo accounts. They firstly agreed to participate in the study by clicking "agree" button after reading the consent form. After then, they authorized our system to download their Weibo records through Sina APIs.[6]

**Behavioral and Linguistic Features**

In this study, we downloaded users' profile and microblogging statuses. According to participants' online activities, we extracted four groups of behavior features as shown in Table 2. For each user's Weibo data, 45 behavioral features[6] and 88 linguistic features[16] were extracted.

- Group profile includes 4 features describing basic information of the user, such as age or gender.
- Group self-presentation is used to identify how the user presents to others. There are 12 features in this group, for example, whether the user set the screen name or update the avatar.
- Group security settings contains 3 features about the privacy settings. This group shows the social security preference of the user, such as whether the comments are available to strangers or not.
- The social networking group, with 26 features, is defined as the online interaction of the user, such as friend count, or follower count of the user.

Table 2. Behavior features in Sina Microblog

| Group | Count | Example |
| --- | --- | --- |
| Profile | 4 | hometown, gender |
| Self-presentation | 12 | screen name, avatar |
| Security settings | 3 | comments available |
| Social networking | 26 | number of friend, follower |

Linguistic features are extracted with TextMind (http://ccpl.psych.ac.cn/textmind/) which is a Chinese language psychological analysis system. TextMind provides easy access to analyze the preferences of different categories in words. Inspired by the dictionary of LIWC2007 and C-LIWC[17,18,19], TextMind is developed based on the characteristics of Simplified Chinese in mainland China It provides an all-in-one solution from automatic Chinese words segmentation to statistical analysis in text words. The dictionary, text and punctuation is compatible with LIWC in simplified Chinese. In this study, the original dictionaries are manually checked by a group of psychology experts, and the extended version of dictionary is developed by a group of experts working on the words from Weibo data.[20] When extracting textual features, TextMind runs with the extended dictionary to calculate 88 microblog status text features.

We used ICTCAS tools[21] for Chinese word segmentation and merge all the Weibo messages of a certain user into a long text as the word bag.[19] Each feature was extracted from the word bag.

In our research, linguistic features can be roughly grouped into five types as shown in Table 3.

- Basic usage describes the word usage situation of Microblogging text, such as the number of words from different parts of speech, sentence or different kinds of punctuations.
- Journey word group includes social, emotion, cognitive, percept and physiological words.
- Personal vocabulary contains family words, working words etc.
- Spoken word is the words usually appear in spoken language, such as echoing words, affixes and pausing words.
- Chinese characteristic group describes the different usage between Chinese and English, such as "ni"

and "nimen", which is the singular form and plural form of the second person in Chinese.

Table 3. Text features in Sina Microblog

| Group | Count | Example |
|---|---|---|
| Basic usage | 35 | Word, comma |
| Journey word | 32 | Emotional word |
| Personal vocabulary | 7 | Vocabulary about work |
| Spoken word | 3 | Affix in speaking |
| Chinese characteristic | 11 | Plural second person |

**Time Slicing Blocks Method**

The online digital records of social media users are traceable, which means former records can be consulted at any time. With this property, we proposed the method of slicing blocks as shown in Figure 1.

At time axis, 13 time points were set which are the first dates of the 13 months from February, 2012 to February, 2013. At each time point, user's behavioral features and linguistic features were extracted from their registration date to each time points. Therefore, for each user, we got 13 sets of features at different time points. At city axis, users with same city information were grouped as the sample set of each city together. Figure 1 is an example of time slicing blocks of Dongguan city at April, 2013. In this research, 13 nodes are set in time axis totally. In city axis, there are 21 cities in Canton altogether. Therefore, $273_{(13*21)}$ slicing blocks of features can be obtained.

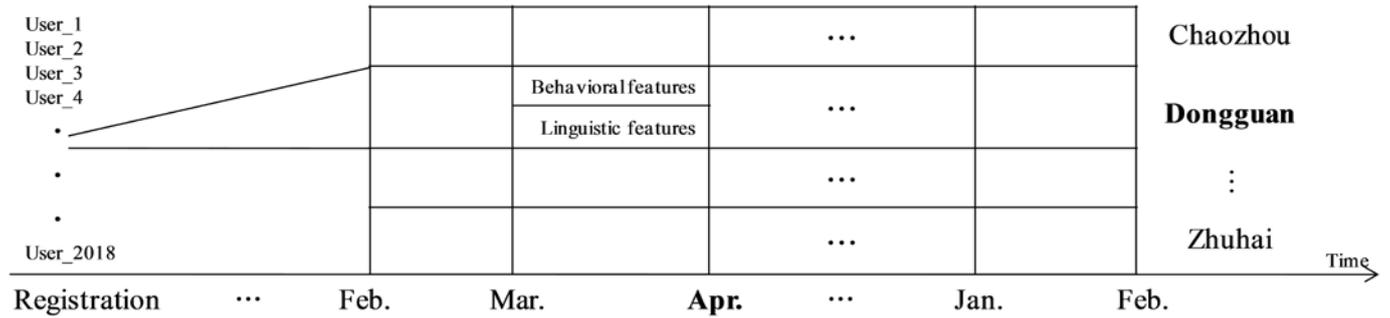

Figure 1. Time slicing blocks of Dongguan at April, 2013.

**Modeling and Prediction**

Since social satisfaction inventory labels (i.e., scores) were continuous, we built the predicting model using regression which is optimized by least square method[22]. We computed Mean Absolute Error (MAE) and Pearson Correlation Coefficient (PCC) between the predicted results and inventory labels to evaluate the performance of the model.

Once the predicting model was built, it can be used to predict user's social satisfaction at any time. In this study, we ran the prediction of social satisfaction from February, 2012 to February, 2013. Since most of the social satisfaction dimensions were economy related, we proposed to compute their relevance with the actual economy indexes in Guangdong to validate the predicted results.

In this paper, all the labels are scored as 5 = strongly satisfy, 4 = satisfy, 3 = neutral, 2 = dissatisfy, 1 = strongly dissatisfy. Table 4 shows the MAE and PCC of different regression models: life satisfaction (LS), income satisfaction (IS), social position satisfaction (SPS), national economy satisfaction (NES), local economy satisfaction (LES), social justice satisfaction (SJS). On average, our model for predicting social satisfaction has an accuracy of 84.50% and a correlation factor of 0.41 using M5P method.

Table 4. The PCC and MAE of the prediction model.

| Dim. | Linear regression | | Pace regression | | M5P | |
|---|---|---|---|---|---|---|
| | PCC | MAE | PCC | MAE | PCC | MAE |
| LS | 0.30 | 0.52 | 0.23 | 0.53 | 0.26 | 0.52 |
| IS | 0.33 | 0.72 | 0.21 | 0.76 | 0.24 | 0.74 |
| SPS | 0.55 | 0.76 | 0.47 | 0.73 | 0.43 | 0.74 |
| NES | 0.54 | 0.74 | 0.41 | 0.71 | 0.54 | 0.62 |
| LES | 0.24 | 0.60 | 0.24 | 0.60 | 0.53 | 0.49 |
| SJS | 0.26 | 0.72 | 0.25 | 0.73 | 0.47 | 0.63 |
| AVE | 0.37 | 0.68 | 0.30 | 0.68 | 0.41 | 0.62 |

**Results**

**Regional Social Satisfaction of Guangdong Province**

We are able to dynamically express the changing tendency of social satisfaction within a specified time interval with our prediction model. To verify the validity of predicted social satisfaction, we calculate the relevance of the predicted results with the specific economical indexes.

For each city in Guangdong, we used time slicing blocks method to extract features, and predicted the social satisfaction of each participant monthly from February 2012 to February 2013.

After then, we acquire social satisfaction variation curve for each city from February, 2012 to February, 2013. Figure 2 shows the social satisfaction curve of Zhongshan, which is a city in Guangdong.

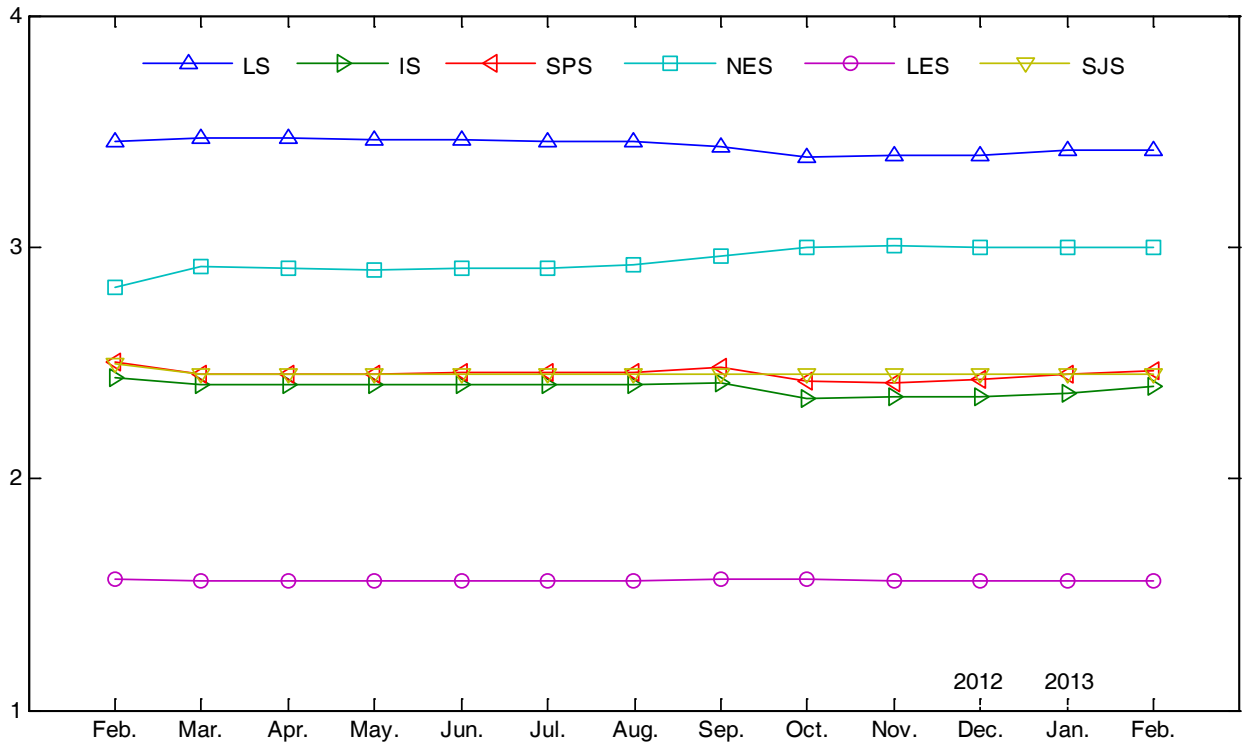

Figure 2. An example of satisfaction variation tendency curve of Zhongshan city.

**Correlation between Social Satisfaction and Economy Indexes**

For each city, we utilize our predicting model to calculate the regional social satisfaction of 21 cities in Guangdong in 2012. As shown in Figure 2, for each social satisfaction dimension, we obtain 13 scores for each month. To acquire the regional social satisfaction of the city in the whole year, we use the median value of all 13 satisfaction scores, which is further used to check its correlation with local economy indexes in 2012.[14] The economy indexes used in this research are GDP, agriculture, retail trades and staff employment. Table 5 shows the PCC between the indexes and the predicted social satisfaction in 2012.

Table 5. PCCs of the actual economy indicators and the social satisfaction of Guangdong province in 2012.

|  | LS | IS | SPS | NES | LES | SJS |
|---|---|---|---|---|---|---|
| 1. Gross Domestic Product (GDP) | 0.06 | 0.10 | -0.26 | 0.05 | 0.15 | -0.12 |
| 2. Growth Rates of GDP | -0.21 | -0.11 | 0.29 | -0.14 | 0.41 | 0.38 |
| 3. Per Capita GDP | 0.15 | 0.13 | 0.26 | -0.08 | 0.66** | 0.25 |
| 4. Gross Output Value of Agriculture | 0.04 | 0.15 | 0.08 | -0.16 | 0.58** | 0.04 |
| 5. Output of Grain | -0.04 | -0.02 | 0.18 | -0.08 | 0.45* | 0.38 |
| -Output of Rice | 0.04 | -0.07 | 0.13 | -0.12 | 0.43 | 0.32 |
| 6. Output of Sugarcane | 0.30 | 0.31 | 0.38 | 0.08 | 0.39 | 0.39 |
| 7. Output of Peanut | 0.11 | 0.08 | 0.32 | -0.15 | 0.49* | 0.39 |
| 8. Output of Vegetable | -0.21 | 0.07 | 0.04 | -0.04 | 0.60** | 0.18 |
| 9. Number of Hogs on Hand at the Year-end | 0.07 | 0.04 | 0.06 | -0.16 | 0.59** | 0.16 |
| 10. Slaughtered Fattened Hogs | 0.08 | 0.02 | 0.00 | -0.19 | 0.58** | 0.09 |
| 11. Output of Pork | 0.07 | 0.02 | -0.01 | -0.18 | 0.58** | 0.09 |
| 12. Total Retail Sales of Consumer Goods | -0.19 | -0.01 | 0.32 | -0.17 | 0.45* | 0.54* |
|     -Total Retail Sales of Wholesale and Retail Trades | -0.19 | -0.01 | 0.33 | -0.15 | 0.44* | 0.54* |
| 13. Number of Fully Employed Staff and Workers | 0.18 | 0.18 | 0.03 | 0.26 | 0.19 | 0.08 |
| 14. Total Wages of Fully Employed Staff and Workers | 0.16 | 0.15 | -0.06 | 0.25 | 0.09 | -0.05 |
| 15. Average Wage of Fully Employed Staff and Workers | 0.03 | -0.09 | 0.15 | -0.11 | 0.51* | 0.33 |

*p<0.05

**p<0.01

## Discussion

In this paper, we built the social satisfaction predicting model based on digital records of Sina Weibo. We downloaded users' microblogging data utilizing APIs and acquired the satisfaction scores by online survey. The regional social satisfaction predicted is found correlated with local economy indexes, which indicates that the predicting model is able to identify social satisfaction from social media. To demonstrate the changing tendency of social satisfaction, time slicing blocks of features are introduced in this research. Social satisfaction is not a trait variable (e.g. big-five personality), but a state variable, which means its value changes along with the external conditions. This is the base of using time slicing blocks method. As shown in Figure 1, time slicing blocks method shows a certain advantage when tracing state variables in psychology.

### Social Satisfaction and Economy Indexes

Using the model, we find that local economy satisfaction (LES) significantly correlates with economy indicators such as per capita GDP ($r = 0.66$), gross output value of agriculture ($r = 0.58$), output of vegetable ($r = 0.60$) and hog related factors ($r = 0.58$). LS, IS, SPS, even NES weakly correlate with economy indicators, compared with LES. It means that local economy satisfaction is reflected by the objective economy indexes closely. Total retail sales of consumer goods, wholesale and retail trades are significantly positive correlated with social justice satisfaction (SJS). An interesting finding is that economic development can rise the degree of local economy satisfaction. Yet, social position satisfaction, life satisfaction and income satisfaction show little relevance with economy development. The possible reason might be most of Microblog participants are youth.[23] In China nowadays, young people focus on study most of their time, and they do not need to face the press of the family life.

From Table 5, social position satisfaction, social justice satisfaction, and life satisfaction are positively correlated with most of the economy indexes weakly. This finding was also reported by Sirgy[26] who conducts a survey at seven major cities in different countries, and proves that economic development contributes positively to life satisfaction. In the previous research on income satisfaction, Clark[24] found that income satisfaction is influenced not only by the detailed income information, but also by the income of the neighbors of the respondents, which is consistent with our results. They researched on well-being and comparisons by dividing the country up into around 9,000 small neighborhoods. They used income information to match demographic and economic satisfaction variables from eight years of Danish ECHP data.

NES (LES) is the satisfaction about the economy development towards the nation (regional city). In Table 5, LES is positively correlated with economy indexes significantly. In contrast, NES is negatively correlated with most of the economy indexes except for the number and wage of fully employed staff and workers. That means for ordinary people, the national economy satisfaction is not decided by regional economy indexes, such as GDP, but affected by their employment situation, job arrangement and benefit.

### Prediction of Social Satisfaction in Social Media

To verify the reliability, we compute the Pearson correlation factors between the predicted social satisfaction and economy indexes in Table 5. Using M5P method, the accuracy of the social satisfaction prediction model is 84.50% with a correlation factor of 0.41 (ranged from 0.24 to 0.54). The results indicate that social media can be used to predict users' social satisfaction fairly well, and the prediction is reliable.

**Future work**

This paper builds the social satisfaction model and verifies the relationship between predicted social satisfaction and economy indicators. The model is training by machine learning algorithm. In the future, we will focus on trying other modeling algorithms, and conducting more user studies to check the validity of the model.

**Acknowledgments**

The authors gratefully acknowledge the generous support from NSFC (61070115), Institute of Psychology (113000C037), Strategic Priority Research Program (XDA06030800) and 100-Talent Project (Y2CX093006) from Chinese Academy of Sciences.